\documentclass[twocolumn]{aastex631}
\usepackage{mathtools} 
\usepackage{csvsimple}
\usepackage{longtable}
\usepackage{hyperref}

\newcommand{\gtcosiris}{GTC/OSIRIS+}
\newcommand{\msun}{\rm{M}_{\odot}}
\newcommand{\rsun}{\rm{R}_{\odot}}
\newcommand{\mjup}{\rm{M}_{\rm Jup}}
\newcommand{\rjup}{\rm{R}_{\rm Jup}}
\newcommand{\teff}{T_{\rm eff}}
\newcommand{\logg}{\log g}
\newcommand{\kms}{km\,s^{-1}}
\newcommand{\angstrom}{\text{\AA}}

\usepackage{xcolor}

\shorttitle{J0404+1112: A Highly-Irradiated WD–BD Benchmark}
\shortauthors{de Wit et al.}

\begin{document}
\title{J0404+1112: A 3-Hour Eclipsing White Dwarf–Brown Dwarf Probing Multiple Atmospheric Regimes}



\author[0000-0003-2415-2191]{J.~de~Wit}
\affiliation{Department of Earth, Atmospheric and Planetary Sciences,
Massachusetts Institute of Technology, Cambridge, MA 02139, USA}

\author[0000-0001-8462-8126]{R.~Alonso}
\affiliation{Instituto de Astrof\'isica de Canarias (IAC),
Calle V\'ia L\'actea s/n, 38200 La Laguna, Tenerife, Spain}
\affiliation{Departamento de Astrof\'isica,
Universidad de La Laguna (ULL),
E--38206 La Laguna, Tenerife, Spain}

\author[0000-0002-3627-1676]{B.~V.~Rackham}
\affiliation{Department of Earth, Atmospheric and Planetary Sciences,
Massachusetts Institute of Technology, Cambridge, MA 02139, USA}
\affiliation{Kavli Institute for Astrophysics and Space Research, Massachusetts Institute of Technology, Cambridge, MA 02139, USA}

\author[0000-0003-2805-8653]{S.~M.~Lederer}
\affiliation{NASA Johnson Space Center,
2101 NASA Parkway, Houston, TX 77058, USA}

\author[0000-0001-9892-2406]{A.~Y.~Burdanov}
\affiliation{Department of Earth, Atmospheric and Planetary Sciences,
Massachusetts Institute of Technology, Cambridge, MA 02139, USA}

\author[0000-0003-2478-0120]{S.L. ~Casewell}
\affiliation{School of Physics and Astronomy,
University of Leicester,
University Road, Leicester LE1 7RH, UK}

\author[0000-0003-2969-6040]{Y. Zhou}
\affiliation{Department of Astronomy, University of Virginia, 530 McCormick Rd., Charlottesville, VA 22904, USA}

\author[0000-0003-0825-4876]{J.~R.~French}
\affiliation{School of Physics and Astronomy,
University of Birmingham,
Edgbaston, Birmingham B15 2TT, UK}

\author[0000-0003-1464-9276]{K.~Barkaoui}
\affiliation{Instituto de Astrof\'isica de Canarias (IAC),
Calle V\'ia L\'actea s/n, 38200 La Laguna, Tenerife, Spain}
\affiliation{Department of Earth, Atmospheric and Planetary Sciences,
Massachusetts Institute of Technology, Cambridge, MA 02139, USA}
\affiliation{Astrobiology Research Unit, Universit\'e de Li\`ege,
All\'ee du 6 ao\^ut 19, 4000 Li\`ege, Belgium}

\author[0009-0008-1188-2025]{M.~Albornoz}
\affiliation{Department of Earth, Atmospheric and Planetary Sciences,
Massachusetts Institute of Technology, Cambridge, MA 02139, USA}
\affiliation{Department of Astronomy,
University of Florida,
Gainesville, FL 32611, USA}

\author[0000-0002-6523-9536]{A.~J.~Burgasser}
\affiliation{Department of Astronomy \& Astrophysics, UC San Diego, La Jolla, CA 92093, USA}

\author[0009-0001-4145-8929]{T.~Cavalier}
\affiliation{Department of Earth, Atmospheric and Planetary Sciences,
Massachusetts Institute of Technology, Cambridge, MA 02139, USA}
\affiliation{Institut Supérieur de l’Aéronautique et de l’Espace (ISAE-SUPAERO), Toulouse, 31055, France}

\author[0000-0002-9355-5165]{B.-O.~Demory}
\affiliation{Centerfor Space and Habitability, University of Bern, Gesellschaftsstrasse
6, 3012, Bern, Switzerland}

\author[0000-0002-7008-6888]{E.~Ducrot}
\affiliation{AIM, CEA, CNRS, Université Paris-Saclay, Université de Paris,
F-91191 Gif-sur-Yvette, France}

\author[0000-0003-1462-7739]{M.~Gillon}
\affiliation{Astrobiology Research Unit,
Universit\'e de Li\`ege,
All\'ee du 6 ao\^ut 19, 4000 Li\`ege, Belgium}

\author[0000-0002-7486-6726]{Y.~Gómez~Maqueo~Chew}
\affiliation{Universidad Nacional Autónoma de México, Instituto de As-
tronomía, AP 70-264, Ciudad de México, 04510, México}

\author[0000-0003-0030-332X]{M.~J.~Hooton}
\affiliation{Cavendish Laboratory, JJ Thomson Avenue, Cambridge CB3 0HE, UK}

\author[0000-0002-5220-609X]{P.~P.~Pedersen}
\affiliation{Cavendish Laboratory, JJ Thomson Avenue, Cambridge CB3 0HE, UK}
\affiliation{Institute for Particle Physics and Astrophysics, ETH Zürich,
Wolfgang-Pauli-Strasse 2, 8093 Zürich, Switzerland}

\author[0000-0002-3012-0316]{D.~Queloz}
\affiliation{Cavendish Laboratory, JJ Thomson Avenue, Cambridge CB3 0HE, UK}
\affiliation{Institute for Particle Physics and Astrophysics, ETH Zürich,
Wolfgang-Pauli-Strasse 2, 8093 Zürich, Switzerland}

\author[0009-0008-2214-5039]{M.~Timmermans}
\affiliation{School of Physics and Astronomy,
University of Birmingham,
Edgbaston, Birmingham B15 2TT, UK}
\affiliation{Astrobiology Research Unit, Université de Liège, 19C Allée du 6 Août, 4000 Liège, Belgium}

\author[0000-0002-5510-8751]{A.~H.~M.~J.~Triaud}
\affiliation{School of Physics and Astronomy,
University of Birmingham,
Edgbaston, Birmingham B15 2TT, UK}

\author[0000-0002-9350-830X]{S.~Zúñiga-Fernández}
\affiliation{Astrobiology Research Unit,
Universit\'e de Li\`ege,
All\'ee du 6 ao\^ut 19, 4000 Li\`ege, Belgium}


\begin{abstract}
White dwarf–brown dwarf (WD+BD) binaries are rare laboratories for probing both substellar survival through post-main-sequence evolution and the physics of strongly irradiated atmospheres. We present time-resolved eclipse photometry and radial-velocity confirmation of J0404+1112, a compact ($P = 2.93$ hr), totally eclipsing WD+BD system initially reported in a Research Note of the AAS.
New high-cadence observations resolve a total eclipse of the WD and, combined with spectroscopy, broadband photometry, and the \textit{Gaia} parallax, yield a revised system architecture. We find a hot DA WD with $T_{\rm eff} \simeq 28{,}000$ K, $\log g \simeq 8.0$, $M_{\rm WD} \simeq 0.6\,\rm{M}_\odot$, and $R_{\rm WD} \simeq 0.015\,\rm{R}_\odot$, orbited by a $\sim40\,\rm{M}_{\rm Jup}$ BD with radius $\sim0.9\,\rm{R}_{\rm Jup}$. These parameters supersede earlier SED-based estimates in which the WD temperature, radius, distance, and extinction were strongly covariant. We break this degeneracy with normalized Balmer-line profile fitting and high-cadence eclipse photometry that constraints the white dwarf radius geometrically.
The total eclipses enable isolation of the BD flux and constrain its nightside emission. 
With its short orbital period and strong irradiation, J0404+1112 provides a uniquely compact, high-sensitivity \textit{JWST} benchmark for probing atmospheric heat redistribution and photochemistry in the temperature regime of the hottest ultra-hot Jupiters, spanning dayside and nightside brightness temperatures of $\sim3{,}600$\,K and $\sim1{,}800$\,K within a single $\sim$3\,hr orbit.
\end{abstract}


\section{Introduction}

Most known exoplanetary systems orbit stars that will evolve through a red-giant
phase and end as white dwarfs (WDs). Substellar companions to WDs therefore encode the outcomes of post-main-sequence evolution and common-envelope processes, with implications for the present-day and future planetary population of the Milky Way \citep[e.g.,][]{VillaverLivio2009}. In addition, irradiated brown dwarfs (BDs) in close WD binaries provide a complementary laboratory to hot and ultra-hot Jupiters \citep{Lothringer2020}, probing many of the same atmospheric processes such as including day--night circulation, vertical mixing, cloud formation, and photochemistry but under distinct irradiation environments---on orbital timescales of only a few hours \citep{Zhou2022}---and with substantially more favorable companion-to-host flux ratios than main-sequence systems. However, WD+BD systems are observationally rare, with only roughly a dozen confirmed examples \citep{French2024}.

Eclipsing WD+BD systems are especially powerful targets within this population.
Their short orbital periods yield large phase modulations, while total eclipses
enable clean isolation of the companion flux and, in favorable cases, direct
measurements of the brown dwarf's nightside emission. Combined with the eclipse
geometry, these systems can also provide direct and relatively model-independent
radius constraints on the brown dwarf companion, which remain difficult to obtain
for isolated substellar objects. As a result, phase-resolved spectroscopy of
WD+BD systems provides unusually direct access to the interplay between
atmospheric processes in tidally locked substellar objects
\citep[e.g.,][]{Moses2011,Kataria2016,TanShowman2020}. Recent \textit{HST} and
\textit{JWST} programs have begun mapping these processes across a small sample
of systems spanning a range of irradiation levels and surface gravities,
enabling the first comparative studies of irradiated substellar atmospheres in
this class \citep{Zhou2022,Amaro2025}.

We recently announced the discovery of J0404+1112 as a fully-eclipsing WD+BD system in a Research Note of the AAS \citep{deWit2025RNAAS}. Based on the initially-available data, the system was interpreted as hosting a low-mass, inflated WD and a low-mass BD companion. Here we present new time-resolved photometric and spectroscopic observations, including high-cadence eclipse data, together with a revised analysis of the WD properties. These data resolve a degeneracy between the WD parameters in the initial
spectral modeling using high-cadence eclipse photometry to geometrically constrain the WD radius, yielding a self-consistent solution corresponding to a standard hot DA white dwarf. DA white dwarfs are classified by their pure Hydrogen atmosphere, exhibiting strong hydrogen Balmer lines, while lacking helium or heavy metals. These Balmer lines, combined with high-cadence photometry, enable a robust solution for constraining the physical properties of compact white dwarf systems.

We present below time-resolved photometric and spectroscopic observations, together with atmospheric modeling, that place J0404+1112 in a distinct region of parameter space among WD+BD binaries, spanning a wide range of atmospheric conditions over its $\sim$3\,hr orbit.

\section{Observations and Data Reduction}

\begin{figure}
\centering
\includegraphics[width=0.48\textwidth]{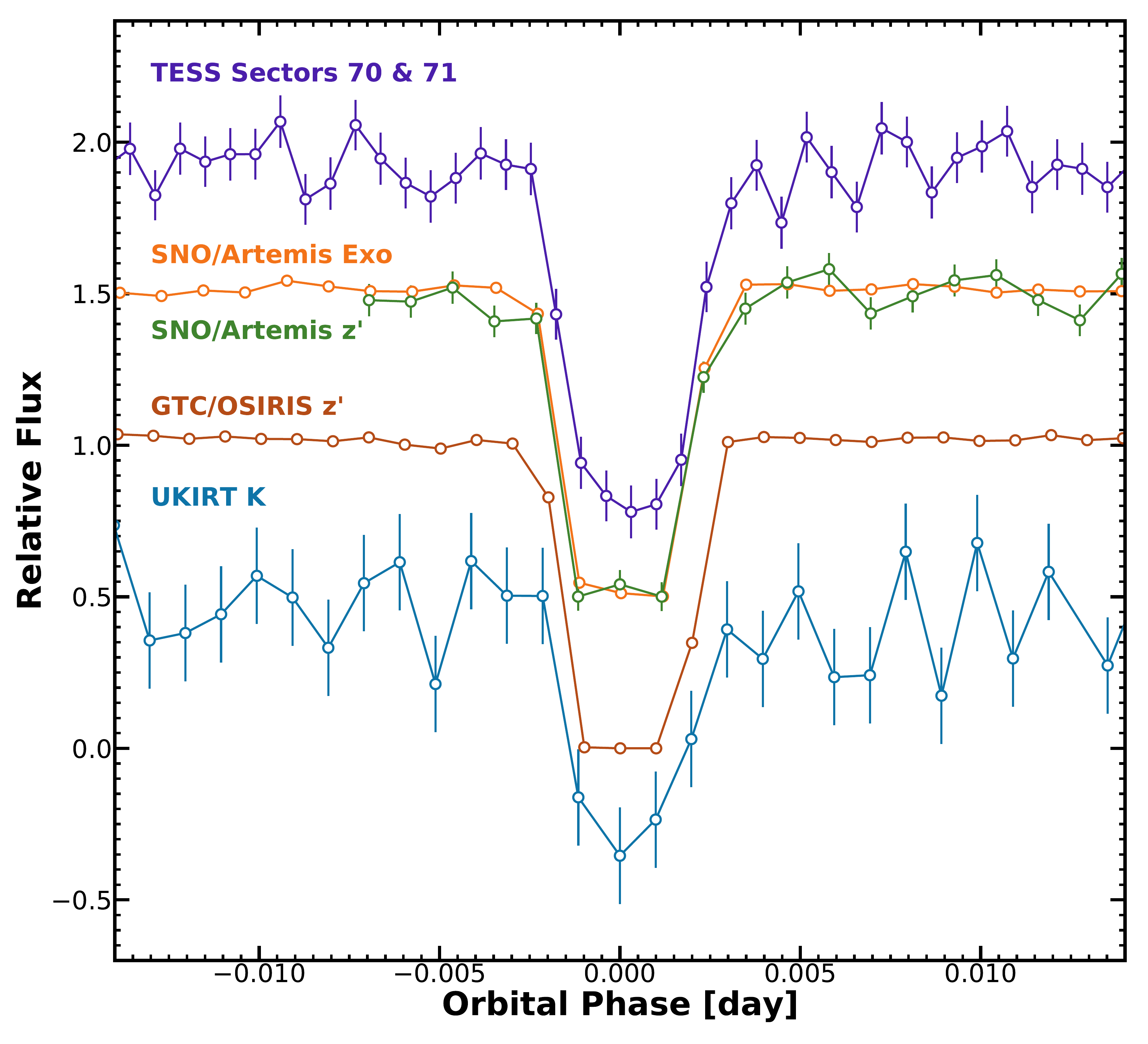}
\caption{\textbf{Low-cadence transit photometry of J0404+1112}.
Phase-folded light curves from TESS, SPECULOOS/Artemis, GTC/OSIRIS, and UKIRT/WFCAMa
time-resolved transit observations, showing a total eclipse consistent with zero in-eclipse
flux. The GTC/OSIRIS provided the definitive time-resolved confirmation of the eclipse shape and
depth and constrain the companion’s $z'$-band flux contribution to $<0.5\%$. The UKIRT K-band measurements provide the first insights into the companion's nightside temperature ($\sim$1,750\,K) via diluted transit depth of 85$\pm$15\%.}
\label{fig:lc}
\end{figure}

\subsection{Transit discovery and ground-based validation}

We searched for short-duration transients in Transiting Exoplanet Survey Satellite (\textit{TESS}; \citealt{Ricker2015}) observations of WDs selected from the Gaia EDR3 WD
catalog \citep{Gentile2021}. Restricting to objects with high-cadence
(\textit{TESS} 2-min) light curves yielded a sample of $\sim37{,}000$ WDs. A
systematic search for short-duration eclipses in these data revealed periodic
events in J0404+1112 (TIC~673440531) at $P\simeq2.93$~hr (see Figure\,\ref{fig:lc}).

\subsubsection{ SPECULOOS-North/Artemis}

Follow-up photometry with the 1-m SPECULOOS-North/Artemis telescope
\citep{Delrez2018,Sebastian2021,Burdanov2022} confirmed that the events are total eclipses
of the WD. As an initial test for wavelength-dependent dilution, we obtained
photometry in several filters: on 2025-08-29 with the custom ``Exo" filter (also known as the ``blue-blocking” filter, which has transmittance $\geq$90\% from 500\,nm to beyond 1000\,nm) and on 2025-08-30 with the Sloan $z^\prime$ filter (transmittance $\geq$90\% from 800\,nm to beyond 1000\,nm). The eclipse
depths are consistent between the two bands within the instrumental precision,
revealing no evidence for color-dependent dilution (Figure~\ref{fig:lc}).

At the precision achieved with this 1-m facility, the non-detection of dilution places an upper limit on the companion's nightside temperature of $\sim2,500$~K (3$\sigma$), mildly favoring a substellar companion. To probe substantially lower temperatures and further constrain the nature of the companion, we obtained follow-up observations with larger aperture facilities.

\subsubsection{GTC/OSIRIS} We obtained time-resolved eclipse photometry of J0404+1112 using OSIRIS \citep{Cepa2000} on the 10.4~m Gran Telescopio Canarias (GTC), in Sloan $z'$ on 2025-09-24.
Despite the increased sensitivity, the eclipse remains consistent with a total occultation of the WD (Fig.\,\ref{fig:lc}), which constrains the companion’s $z'$-band flux contribution to $<0.5\%$ corresponding to a nightside temperature below $\sim 2000\,$K. Additional GTC observations obtained at higher cadence with HiPERCAM \citep{Dhillon2021} are introduced in Section\,\ref{sec:phot}.

\subsubsection{UKIRT/WFCAM}

We observed an eclipse of J0404+1112 on 2025-09-30 with the WFCAM near-infrared
imager on the 3.8-m United Kingdom Infra-Red Telescope (UKIRT) in the broad-band
$K$ filter (2.0--2.5~$\mu$m; \citealt{Hodgkin2009_WFCAM_phot}). 
WFCAM consists of four $2048\times2048$ Rockwell Hawaii-II detectors with a field
of view of $13.65\times13.65$~arcmin$^2$ each and an image scale of
0.4~arcsec~pixel$^{-1}$ \citep{Casali2007_WFCAM}. We placed J0404+1112 on
quadrant 3 (array ID ``RSC:H2:76''), which is the least affected by detector
artifacts, in order to optimize the extraction of a high-noise, red-band light
curve for this faint target ($K=18.4\pm0.1$). This quadrant has a gain of
4.5~e$^{-}$/ADU, a read-out noise of 25~e$^{-}$, and a dark current of
9~e$^{-}$~s$^{-1}$~pixel$^{-1}$.

The data were pre-processed by the Cambridge Astronomy Survey Unit (CASU),
including dark subtraction, flat-fielding, gain correction, sky subtraction, and
decurtaining \citep{Casali2007_WFCAM}. No linearity correction was applied, as
the detector response is linear to better than 1\% up to $\sim30{,}000$ counts,
well above the maximum counts ($\sim15{,}000$) recorded for J0404+1112.

We performed differential aperture photometry on the sky-subtracted images using
the \texttt{prose} package \citep{Prose_Garcia_2002} following the method of
\citet{Broeg_2005}. The resulting light curve is shown in
Figure~\ref{fig:lc} and exhibits an out-of-transit scatter of 15\%. The transit is
consistent with a 85$\pm$15\% occultation of the WD, which translates to a nightside temperature of $\sim$1,750~K for the companion when considering the WD's properties introduced in Section\,\ref{sec:atm}.


\subsection{Spectroscopy and Radial Velocities}
\label{sec:spec}
 
Time-resolved optical spectroscopy of J0404$+$1112 was obtained with
\gtcosiris{} \citep{Cepa2000} on the night of 2025-09-27.  We used the R2000B Volume Phase
Holographic (VPH) grism with a 0.6\farcs ~long slit. This configuration provides a wavelength coverage of
$3950$--$5700$~\angstrom{} at a spectral resolution of
$R \approx 2170$. The detector was read out with $2\times2$ on-chip binning at 233~kHz
(gain~$=1.9$~e$^{-}$\,ADU$^{-1}$; readout noise~$= 4.3$~e$^{-}$~rms).
 
We obtained 5 consecutive spectra of 300~s at each of the quadratures.  Arc lamp frames (Xe$+$Ne$+$HgAr) and flat-field frames were acquired
during evening twilight. 

\begin{figure}
\centering
\includegraphics[width=1.02\columnwidth]{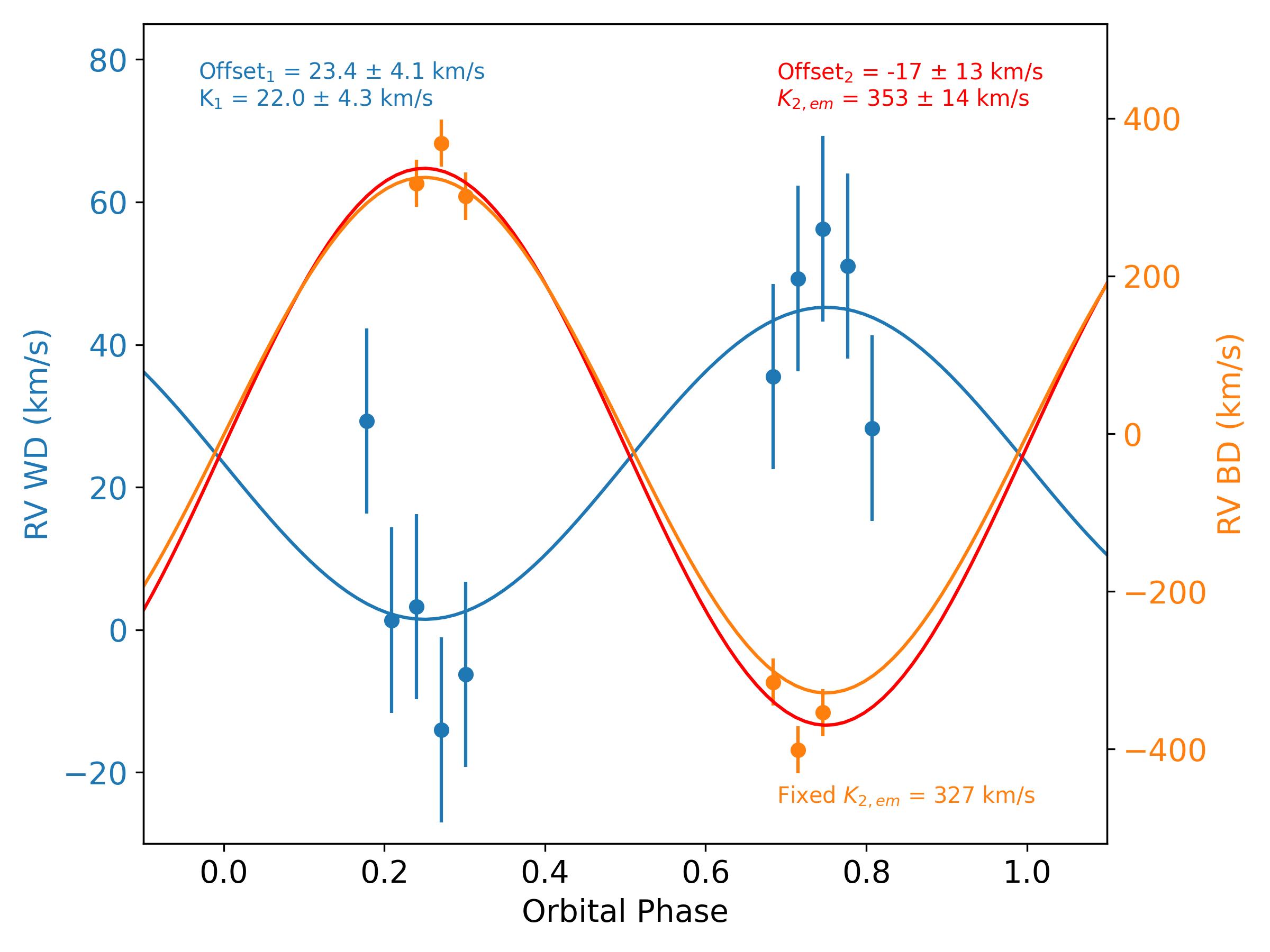}
\caption{Radial velocities of J0404+1112 from GTC spectroscopy. Blue data, lines, and axis represent the RVs of the WD component, while the orange and red are for the brown-dwarf. The best-fit circular orbital solution yields a WD semi-amplitude $K_{\rm WD} = 22.0 \pm 4.3$~km~s$^{-1}$
when the ephemeris is fixed to the transit timing, implying a low-mass brown dwarf companion. The fit to the emission line in the H$\beta$ region (red line) is in reasonable agreement with the solution using the mass function and the spectroscopic mass estimate for the WD (orange line).}
\label{fig:rv}
\end{figure}

\begin{figure*}
\centering
\includegraphics[width=\textwidth]{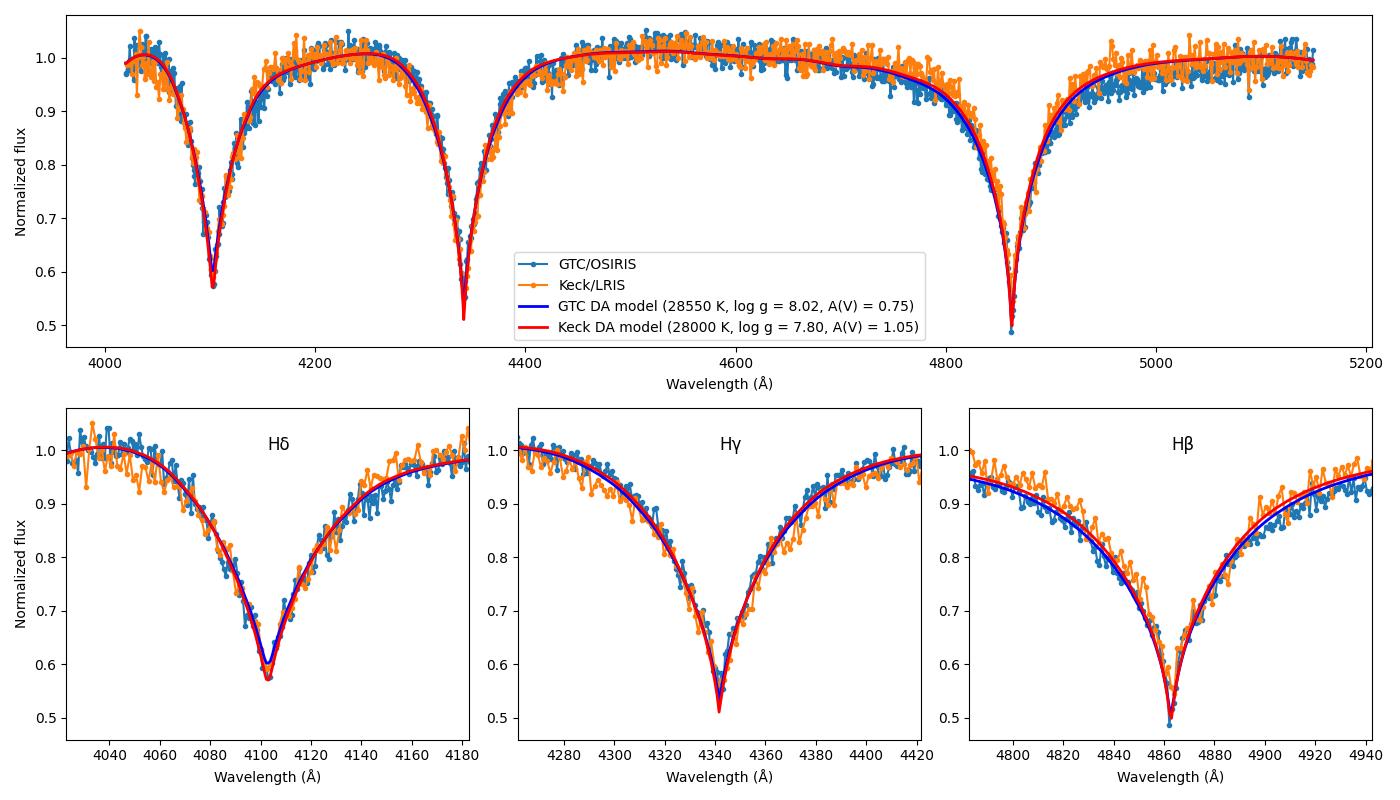}
\caption{Keck/LRIS (orange) and GTC-averaged (blue) spectra of J0404+1112, fit with DA white-dwarf atmosphere models, including reddening and a Gaia parallax prior. The best-fit models and parameters for each spectrum are displayed as solid lines and at the legend, respectively.}
\label{fig:spec}
\end{figure*}

The data were reduced using \texttt{PypeIt} \citep{Prochaska2020}. The radial velocity for each spectrum was obtained through a normalization of the spectra, followed by a fit of the H$\beta$, H$\gamma$ and H$\delta$ lines. Each line was fit using a two-gaussian model, the widths of each gaussian fixed to the fit of the 10-exposures combined spectrum. The free parameters were the amplitudes of the gaussians and a vertical offset for each of the lines, plus a common radial velocity Doppler shift. Some of the spectra show emission features from the irradiated secondary, and we masked these regions before obtaining the RV for the estimation of the $v_{\rm WD}$, or fitted a narrow Gaussian to the emission component in H$\beta$ line for the estimation of $v_{\rm BD,em}$. A circular orbital fit to the RVs yields a WD semi-amplitude $K_{\rm WD}=22.0\pm4.3~\mathrm{km~s^{-1}}$ and $K_{\rm BD,em}=353\pm14~\mathrm{km~s^{-1}}$, with a systemic velocity of $\gamma_1 = 23.4 \pm 4.1~\mathrm{km~s^{-1}}$ and $\gamma_2 =  -17\pm 13~\mathrm{km~s^{-1}}$ (Figure~\ref{fig:rv}). The difference in systemic velocities was used to determine the gravitational redshift as $V_z= 40\pm 14~\mathrm{km~s^{-1}}$. 

 The observed velocity amplitude, in conjunction with the orbital period $P$, yields a
mass function
\begin{equation}
  \label{eq:massfunc}
  f(M) = \frac{M_{\rm BD}^{3}\sin^{3}i}{\left(M_{\rm WD} + M_{\rm BD}\right)^{2}}
       = \frac{K_{\rm WD}^{3} P}{2\pi G}
       = (0.536 \pm 0.051)~\msun.
\end{equation}
 
Combining Equation~(\ref{eq:massfunc}) with the white dwarf mass derived
from atmospheric model fitting (Section~\ref{sec:atm}) and the orbital
inclination from eclipse modeling (Section~\ref{sec:phot}) yields a brown
dwarf mass of $M_{\rm BD} = 39 \pm 8~\mjup$ (Table~\ref{tab:sysparams}). This mass would imply a $K_{\rm BD,em}=327\pm3~\mathrm{km~s^{-1}}$ when considering a redistribution factor $f$ of 0.5 \citep{Parsons12}, which is within 2$\sigma$ from our previous estimation using the few spectra for which this velocity could be measured.

\subsection{White Dwarf Atmospheric Parameters}
\label{sec:atm}
 
The WD effective temperature and surface gravity were determined by fitting the H$\beta$ through H$\delta$ Balmer absorption line profiles in the time-averaged out-of-eclipse spectrum with the grid of synthetic DA white dwarf model spectra of \citet{koester2010}. The models were bi-spline interpolated, multiplied by the instrumental throughput, reddened using the \citet{fitz2007} extinction law and normalized using the same spectral regions and low-order polynomial as the observed spectrum. We used $\chi^2$-minimization and the \textit{emcee} sampler \citep{foreman2013} to explore the parameter space. To account for systematics in the extraction of the spectra, we applied the same methodology to a spectrum obtained with LRIS at Keck, obtained in 2021-11-01 (PID C289; PI: van Roestel). The spectra and best-fit models are displayed in Fig.~\ref{fig:spec}. We report as final values the average of the parameters determined with each spectrum. We obtain $\teff(\rm WD) = 28300 \pm 250$~K and $\logg = 7.9 \pm
0.1$, corresponding to $M_{\rm WD} = 0.59 \pm 0.05~\msun{}$ via the theoretical evolutionary sequences of \citet{bedard2020}. To make the model spectrum agree with the reported magnitudes and distance (Table~\ref{tab:sysparams}), an extinction A$_V \sim$0.9 is needed, which is in reasonable agreement with the extinction maps of \cite{schleg1998} (A$_V$=0.918) and \cite{schla2011} (A$_V$=0.7895)---see Table~\ref{tab:sysparams}. 
A comparison between this analysis and the initial characterization of \citet{deWit2025RNAAS} is provided in Appendix~\ref{appendix:comparison}.

\subsection{GTC high-cadence transit photometry}
\label{sec:phot}

\begin{figure}[ht]
\centering
\includegraphics[width=\columnwidth]{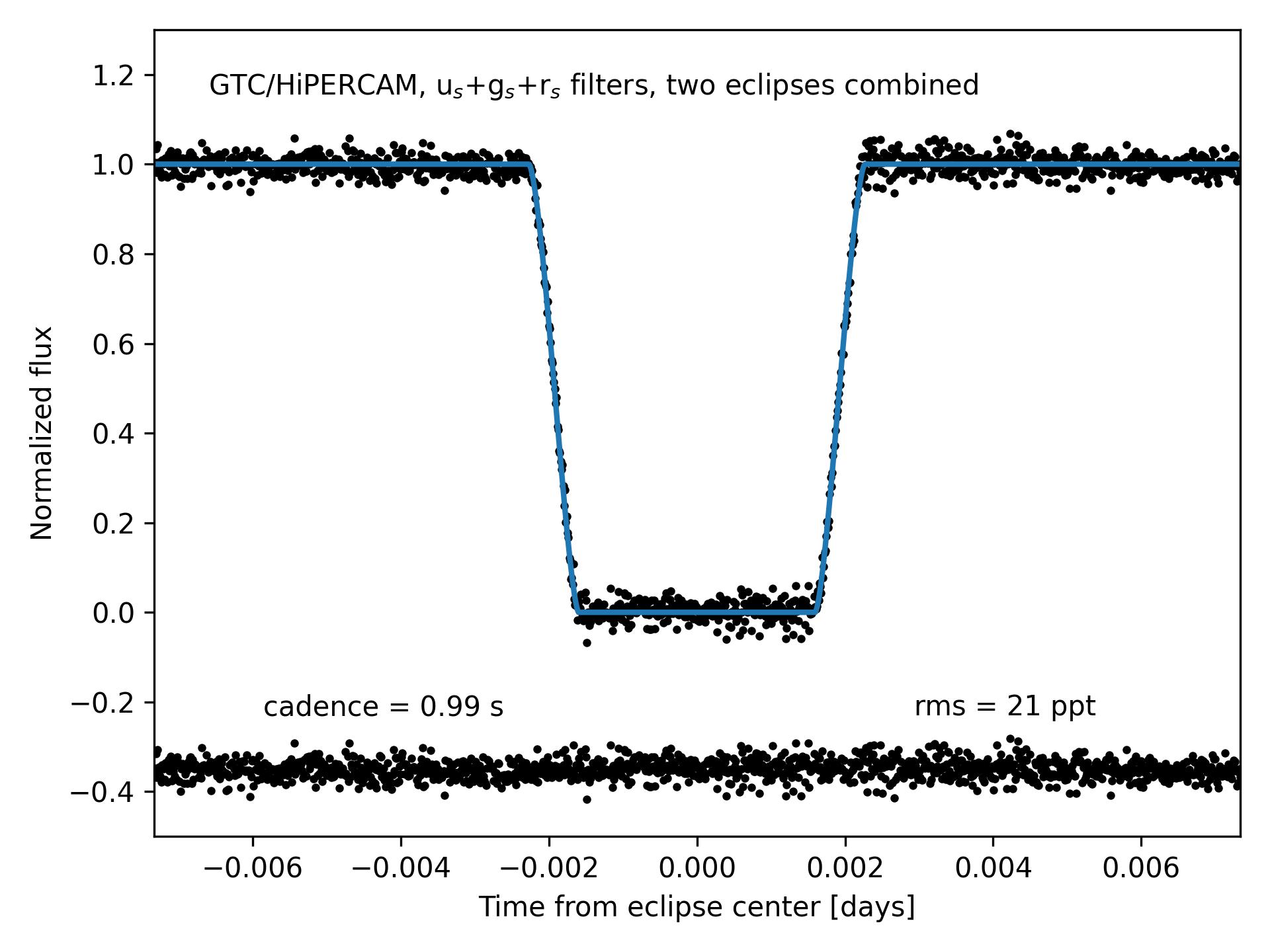}
\caption{The GTC/HiPERCAM light curve of two combined eclipses, summing the light from three HiPERCAM channels. The best-fit eclipse model is shown with a blue line, and the residuals of this model are displayed at the bottom part of the plot.}
\label{fig:gtc_hiper}
\end{figure}

 Two additional eclipses were observed with GTC, using the 5-simultaneous channel instrument HiPERCAM \citep{Dhillon2021} on the nights of 2026-01-15 and 2026-01-16. The exposure time and cadence was of 0.568~s in all the $g_s,r_s,i_s,z_s$ bands, and 1.136~s in the $u_s$ band. Differential photometry was extracted using the HiPERCAM pipeline \citep{Dhillon2021}. We combined the two HiPERCAM observations in the $u_s,g_s,r_s$ bands, and binned to a 1~s cadence. The resulting curve was analyzed using \texttt{exoplanet} \citep{foreman2021}, using as prior the $M_{\rm WD}$ determined in the previous section. The fitted parameters were the mean value out of eclipse, the eclipse center, the radius of both components, the impact parameter, and the quadratic limb darkening coefficients of the WD, using the \cite{kipping2013} parameterization. The results of the fit are listed in Table~\ref{tab:sysparams}, and the folded combined curve with the best fit model and residuals in Figure~\ref{fig:gtc_hiper}.
 This geometric constraint on the WD radius is central to distinguishing the standard-radius WD solution from the inflated WD solution allowed by the preliminary SED-based analysis \citep{deWit2025RNAAS}.
The spectroscopically inferred mass and eclipse-derived radius for the white dwarf imply a gravitational redshift of $V_z= 25\pm 2~\mathrm{km~s^{-1}}$, within the 1$\sigma$ values of the redshift measured from the systemic velocities of the two components.

\section{Discussion}

\begin{figure*}
    \centering
    \includegraphics[width=\textwidth]{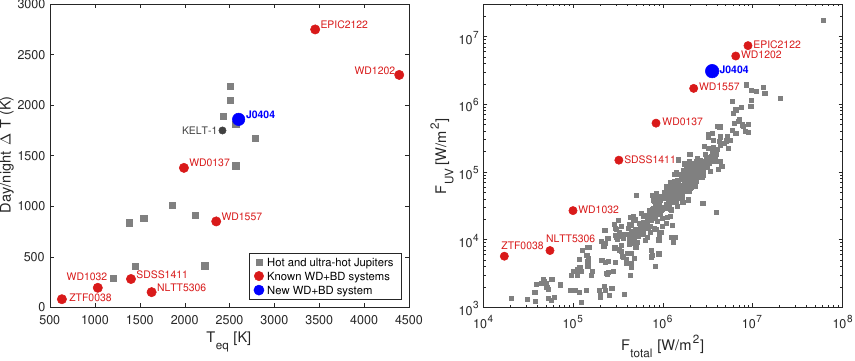}
    \caption{
\textbf{J0404+1112 in the context of irradiated brown dwarfs and (ultra-)hot Jupiters.}
Left: day--night temperature contrast as a function of equilibrium temperature for hot and ultra-hot Jupiters and known white-dwarf--brown-dwarf (WD+BD) systems. Right: UV irradiation flux versus total irradiation flux for the same systems. J0404+1112 lies near the equilibrium-temperature regime of the hottest ultra-hot Jupiters, around $T_{\rm eq}\sim2500$~K, while experiencing a distinct WD irradiation environment, making it a uniquely-suited benchmark for testing atmospheric heat redistribution and photochemistry at ultra-hot-Jupiter temperatures. The canonical irradiated brown dwarf KELT-1b is shown for comparison; however, the compact geometry and total eclipses of J0404+1112 enable substantially cleaner and more efficient atmospheric characterization.
}
    \label{fig:wdbd_comparison}
\end{figure*}

J0404+1112 is a short-period, totally eclipsing WD+BD system whose architecture is now securely established through a joint analysis of time-resolved photometry, spectroscopy, and atmospheric modeling. In contrast to the initial interpretation presented in \citet{deWit2025RNAAS}, the revised solution corresponds to a standard hot DA white dwarf orbited by a substellar companion. As discussed in Appendix~\ref{appendix:comparison}, the initial low-gravity solution arose from a continuum-dominated SED fit in which the WD temperature, radius, distance, and extinction were strongly covariant. The present Balmer-line analysis and high-cadence eclipse photometry break this degeneracy and favor a hotter, higher-gravity WD.

In this configuration, J0404+1112 occupies a particularly valuable region of parameter space among irradiated substellar companions. Its short orbital period and totally eclipsing geometry enable efficient phase-resolved characterization, while the companion spans a wide range of atmospheric conditions over a single $\sim$3\,hr orbit, from a cooler nightside to a strongly irradiated dayside. This makes J0404+1112 a favorable target for probing day--night temperature contrasts, circulation efficiency, and disequilibrium chemistry in irradiated substellar atmospheres \citep{Moses2011,Kataria2016,TanShowman2020}, as well as two- and three-dimensional studies via eclipse mapping \citep{Majeau2012,deWit2012,Challener2022}.

Figure~\ref{fig:wdbd_comparison} places J0404+1112 in the broader context of hot and ultra-hot Jupiters and the growing sample of irradiated WD+BD systems. The hot-Jupiter temperature contrasts are drawn from the Spitzer phase-curve analysis of \citet{Bell2021}, while the WD+BD temperature contrasts are taken from the HST phase-curve series of \citet{Zhou2022}, \citet{Lew2022}, \citet{Amaro2023}, \citet{Amaro2024}, \citet{French2024}, and \citet{Amaro2025}. J0404+1112 lies near the equilibrium-temperature regime of the hottest hot and ultra-hot Jupiters, around $T_{\rm eq}\sim2500$~K, while exhibiting a large expected day--night temperature contrast. It is placed in a sparse region of parameter space, bridging the gap between moderately and highly irradiated WD+BD binaries. It therefore probes thermal regimes comparable to the most strongly irradiated giant planets, but in a compact, totally eclipsing system where both hemispheres can be isolated over a single short orbit.

In hot and ultra-hot Jupiters, dayside temperatures of several thousand kelvin can drive molecular dissociation, ionization, thermal inversions, and enhanced photochemistry, while nightside cooling allows recombination, molecule-dominated chemistry, and potentially cloud formation \citep[e.g.,][]{Lothringer2020}. Simple equilibrium considerations yield a characteristic dayside temperature of $T_{\rm day}\approx2600$~K, while assuming a Bond albedo of 0.5 and partial heat redistribution gives a nightside temperature of $T_{\rm night}\approx1850$~K (Table~\ref{tab:sysparams}). The nightside estimate is consistent with the temperature inferred from the UKIRT photometry ($\sim1750$~K), providing an independent check on the irradiation regime of the system. The resulting contrast places J0404+1112 among the most informative known WD+BD systems for testing atmospheric heat redistribution across the hot- to ultra-hot-Jupiter transition.

At the same time, J0404+1112 is not simply an analog of a hot Jupiter. Its WD provides a distinct irradiation spectrum, including a comparatively strong UV component for a given bolometric irradiation level. In Figure~\ref{fig:wdbd_comparison}, the total irradiation flux is estimated from the Stefan--Boltzmann law, while the UV flux is computed by integrating a blackbody spectrum shortward of $3000$~\AA. This places WD+BD systems in a complementary irradiation environment to main-sequence hot Jupiters, providing a valuable lever arm for disentangling the effects of equilibrium temperature, irradiation spectrum, surface gravity, and atmospheric composition on heat redistribution and photochemistry.

J0404+1112 also shares key atmospheric parameters with the canonical irradiated brown dwarf KELT-1b \citep{Siverd2012}, a $\sim27\,M_{\rm Jup}$ companion orbiting an F star every 1.27\,days with an equilibrium temperature of $\sim2400$~K \citep{Beatty2019}. Despite broadly similar irradiation levels and thermal regimes, WD+BD systems offer several important observational advantages over the few dozens BDs transiting main-sequence stars \citep[e.g.,][]{Henderson2024}. The compact size and infrared faintness of the white-dwarf host, combined with the total eclipses observed in J0404+1112, enable a much cleaner isolation of the brown dwarf flux than is typically achievable for systems such as KELT-1b. This geometry also enables higher-sensitivity phase-resolved atmospheric characterization and efficient full-orbit spectroscopy. In addition, the short orbital period allows a complete phase curve to be obtained within a single $\sim$3\,hr orbit. KELT-1b exhibits an equatorial jet which redistributes heat from the irradiated dayside to the non-irradiated nightside, a feature that is not seen in the atmospheres of WD+BD binaries with similar day-night temperature contrasts \citep[e.g.,][]{Zhou2022,French2024}. With such similar parameters, J0404+1112 therefore provides an important comparative benchmark for investigating how the irradiation spectrum, surface gravity, and rotation shape atmospheric circulation and heat redistribution in strongly irradiated substellar atmospheres.

More broadly, J0404+1112 adds to the small but growing sample of eclipsing WD+BD systems enabling comparative studies of irradiated atmospheres under distinct irradiation environments and surface gravities \citep[e.g.,][]{Zhou2022,French2024}. Its geometry and timescale provide access to both the dayside and nightside emission of the companion, offering direct constraints on energy transport and atmospheric structure. Measurements of the nightside flux further inform substellar structure and cooling models \citep{Casewell2020}. Recent \textit{JWST} programs have begun extending such comparative studies into the spectroscopic regime, and J0404+1112 is especially well suited for this next step because its short period, total eclipses, and large expected temperature contrast make it possible to obtain high-sensitivity phase-resolved spectra efficiently.

Finally, this system highlights an important methodological point for the interpretation of survey-era eclipsing white dwarfs. Low-cadence eclipse photometry alone constrains a degenerate locus in the white-dwarf mass--radius plane, while spectroscopic solutions can admit covariant solutions in temperature, radius, distance, and extinction. For J0404+1112, this produced an initially plausible low-mass, inflated WD solution. A revised analysis based on normalized Balmer-line profiles, together with high-cadence eclipse photometry that resolves ingress, totality, and egress---yielding a geometric constraint on the WD radius---robustly recovers the physical configuration of the system.

\section{Conclusion}

J0404+1112 was initially interpreted as a low-mass, inflated WD system due to degeneracies between WD atmospheric parameters and extinction in spectral modeling. This case illustrates the limitations of survey-era analyses based on low-cadence eclipse photometry and broadband SED fitting alone, and highlights the importance of combining spectroscopic characterization with high-cadence eclipse photometry to robustly constrain the physical properties of compact white-dwarf systems.

With this degeneracy resolved, J0404+1112 is established as a short-period, totally eclipsing WD+BD system consisting of a hot DA white dwarf orbited by a substellar companion. Its geometry enables efficient phase-resolved access to both the dayside and nightside emission of the companion, while its $\sim$3\,hr orbit spans atmospheric conditions from ultra-hot ($\sim$3,600K) daysides to relatively cool ($\sim$1,800K) nightsides within a single object.

J0404+1112 is especially valuable because it lies in the equilibrium-temperature regime of the hottest hot and ultra-hot Jupiters, near $T_{\rm eq}\sim2500$~K, while offering the observational advantages of a compact, totally eclipsing WD+BD system. It therefore provides a uniquely suited benchmark for testing atmospheric heat redistribution, photochemistry, and cloud formation under ultra-hot-Jupiter-like temperatures, but in a distinct white-dwarf irradiation environment. This makes J0404+1112 an exceptional target for high-sensitivity \textit{JWST} phase-resolved spectroscopy and a key addition to the comparative study of irradiated substellar atmospheres.

\section*{Acknowledgements}
The authors thank A.~Boogert for contributing to the SpeX observations. J.d.W. and MIT gratefully acknowledge financial support from the Heising-Simons Foundation, Dr. and Mrs. Colin Masson and Dr. Peter A. Gilman for Artemis, the first telescope
of the SPECULOOS network situated in Tenerife, Spain. M. Gillon is FNRS Research Director. He and
ULiege thank the Wallonian Region for its contribution to the funding of the SPECULOOSNorth/Artemis telescope. R.A. acknowledges support from the Spanish Research Agency of the Ministry of Science, Innovation and Universities (AEI-MICIU) under grant PID2023-149439NB-C41. Based on observations made with the GTC telescope, in the Spanish
Observatorio del Roque de los Muchachos of the Instituto de Astrofısica de Canarias, under Director’s
Discretionary Time. Observations made with the Wide-Field Camera (WFCam) on the UKIRT
telescope were granted through Director’s Discretionary Time. UKIRT is owned by the University of
Hawaii (UH) and operated by the UH Institute for Astronomy. Authors would like to thank Mike Irwin from CASU for his help with access to the UKIRT data at WSA and for data calibration; Director of Technical Operations Dr. Tom Kerr and Senior Support Astronomer Dr. Watson Varricatt; Telescope Support Specialist Felipe Olivares Estay, Telescope Operators Alani Miyoko, Alexander C. Thompson for scheduling and performing observations with the UKIRT.
This material is based upon work supported by the National Aeronautics and Space Administration under Agreement No.\ 80NSSC21K0593 for the program ``Alien Earths.''
The results reported herein benefited from collaborations and/or information exchange within NASA's Nexus for Exoplanet System Science (NExSS) research coordination network sponsored by NASA's Science Mission Directorate.
Funding for KB was provided by the European Union (ERC AdG SUBSTELLAR, GA 101054354).

\section*{Data Availability}
The data underlying this article will be shared on reasonable request to the corresponding author.

\bibliography{bibliography}{}
\bibliographystyle{aasjournal}


\appendix
\section{Revised Constraints on the White Dwarf Parameters}
\label{appendix:comparison}

J0404+1112 was first reported in a Research Note by \citet{deWit2025RNAAS} as a fully eclipsing WD+BD system with an unusually low-mass, inflated WD primary.
That preliminary interpretation was based on a joint fit to the broadband SED, available optical/NIR spectra, and the Gaia parallax, allowing the WD effective temperature, surface gravity, radius, distance, and extinction to vary simultaneously.
While this analysis correctly identified the system as a compact eclipsing WD with a substellar companion, the inferred parameters were affected by strong covariances in this multi-dimensional parameter space.

In particular, the global SED fit admitted a solution in which a cooler, lower-gravity, larger radius WD, combined with relatively high extinction, reproduced the observed continuum shape and absolute flux scale.
As the available eclipse data did not yet resolve the event ingress and egress, there was no independent geometric radius constraint capable of breaking this degeneracy.
The resulting low-gravity solution therefore represented a plausible preliminary interpretation of the discovery data, but not a unique physical solution.

The present analysis supersedes that initial characterization in two ways.
First, we determine the WD atmospheric parameters from the normalized Balmer line profiles, applying the same local continuum normalization to both the observed spectra and the DA atmosphere models.
This approach is less sensitive to broadband flux-calibration uncertainties and to the covariance between extinction and the global SED shape.
The Balmer line morphology favors a hotter, higher-gravity DA white dwarf, rather than the cooler, lower-gravity solution inferred from the preliminary SED analysis.

Second, the new high-cadence HiPERCAM eclipse photometry resolves the ingress, totality, and egress of the WD eclipse.
The eclipse morphology provides a geometric constraint on the relative radii and orbital configuration, yielding a WD radius consistent with a standard-mass DA WD.
This independent constraint disfavors the inflated WD interpretation and supports the revised solution derived from the Balmer line profiles.

We therefore attribute the difference between the Research Note parameters and those reported here primarily to the replacement of a continuum-dominated SED inference with a joint interpretation based on Balmer-line spectroscopy and high-cadence eclipse geometry.
This updated analysis breaks the temperature--radius--extinction degeneracy that affected the preliminary characterization.


\begin{deluxetable*}{llc}
\tablecaption{System Parameters for J0404$+$1112 (Gaia DR3 ID 3303804808200625280)
              \label{tab:sysparams}}
\tablewidth{0pt}
\tabletypesize{\small}
\tablehead{
  \colhead{Parameter} &
  \colhead{Value} &
  \colhead{Unit}
}
\startdata
\sidehead{Astrometry and Photometry}
Right ascension (R.A., J2000)\tablenotemark{a}
        & 04:04:52.9             & h m s                      \\
Declination (Decl., J2000)\tablenotemark{a}
        & +11:12:59.2         & $\degr$ $\arcmin$ $\arcsec$ \\
Gaia parallax\tablenotemark{a}
        & $\varpi = 2.528 \pm 0.214$ & mas                     \\
Distance\tablenotemark{b}
        & $d = 404^{+39}_{-32}$   & pc                         \\[4pt]
$G$\tablenotemark{a}
        & $18.602 \pm 0.003$        & mag (Vega)                   \\
$r$ (SDSS)\tablenotemark{a}
        & $18.7 \pm 0.1$        & mag (AB)                   \\
$K$ (UKIDSS)\tablenotemark{d}
        & $18.4 \pm 0.3$        & mag (Vega)                 \\
$W_1$ (WISE)\tablenotemark{d}
        & $18.3 \pm 0.2$        & mag (Vega)                 \\[4pt]
\sidehead{Orbital Parameters}
Orbital period $P_{\rm orb}$
        & $0.1222940 \pm 0.0000002$ & d                           \\
Reference epoch $T_0$\tablenotemark{e}
        & $2460918.6892168 \pm 0.0000010$ & BJD$_{\rm TDB}$ \\
Orbital inclination $i$
        & $89.6 \pm 0.3 $   & $\degr$                    \\
\sidehead{White Dwarf Parameters}
Effective temperature $\teff(\rm WD)$
        & $28300 \pm 250$         & K                          \\
Surface gravity $\logg(\rm WD)$
        & $7.9 \pm 0.1$         & (cgs)                      \\
Mass $M_{\rm WD}$
        & $0.59 \pm 0.05$         & $\msun$                      \\
Radius $R_{\rm WD}$, DA models
        & $0.013 \pm 0.001$     & $\rsun$                      \\
Radius $R_{\rm WD}$, measured
        & $0.0152 \pm 0.0005$         & $\rsun$           \\
WD RV semi-amplitude $K_{\rm WD}$
        & $21.9 \pm 4.4$          & $\kms$                       \\
Systemic velocity $\gamma$
        & $23.2 \pm 4.2$          & $\kms$                       \\
Gravitational redshift $v_{\rm grav}$
        & $25 \pm 2$          & $\kms$                       \\[4pt]
\sidehead{Brown Dwarf Parameters}
Spectral type SpT$_{\rm BD}$
        & M7--L3                  & ...                    \\
Mass $M_{\rm BD}$
        & $39 \pm 8$          & $\mjup$                      \\
Radius $R_{\rm BD}$
        & $0.089 \pm 0.002$     & $\rsun$                      \\
Radius $R_{\rm BD}$
        & $0.87 \pm 0.02$         & $\rjup$                      \\
Surface gravity $\logg(\rm BD)$
        & $5.1 \pm 0.1$         & (cgs)                      \\
BD RV semi-amplitude $K_{\rm BD}$\tablenotemark{h}
        & $346 \pm 4$        & $\kms$                       \\
BD RV semi-amplitude (emission) $K_{\rm BD,em}$\tablenotemark{i}
        & $327 \pm 3$        & $\kms$                       \\
BD equilibrium temperature $T_{\rm eq}$\tablenotemark{j}
        & $2620 \pm 60$          & K                          \\
BD night-side temperature $T_{\rm n}$\tablenotemark{k}
        & $1852 \pm 43$          & K                          \\[4pt]
\sidehead{Derived Quantities}
Mass ratio $q = M_{\rm BD}/M_{\rm WD}$
        & $0.063 \pm 0.013$       & ...                 \\
Mass function $f(M)$
        & $0.523 \pm 0.051$ & $\msun$            \\
\enddata

\tablenotetext{a}{From Gaia DR3 \citep{gaiadr3}.}
\tablenotetext{b}{From Bayesian inference using an exponential decreasing space density prior with a scale length of 1.35 kpc.}
\tablenotetext{e}{$T_0$ is defined as the mid-point of primary eclipse
  (inferior conjunction of the brown dwarf; white dwarf behind the brown
  dwarf). Phase $\phi = 0$ corresponds to $T_0$.}

\tablenotetext{h}{$K_{\rm BD}$ derived from the mass function, white dwarf mass,
  and orbital inclination: $K_{\rm BD} = K_{\rm WD}(M_{\rm WD}/M_{\rm BD})$.} 
\tablenotetext{i}{$K_{\rm BD,em}$ is the RV semi-amplitude assuming a
  centre-of-mass velocity due to the displacement of the emitting region
  toward the sub-stellar point with a $f$=0.5 \citep{Parsons12}.}
\tablenotetext{j}{Assuming isotropic redistribution of the incident flux and zero Bond albedo: $T_{\rm eq}=T_{\rm eff,WD}\sqrt{\frac{R_{\rm WD}}{2a}}$}
\tablenotetext{k}{Assuming Bond albedo = 0.5, $\epsilon$=0.5: $T_{\rm n}=T_{\rm eq}(1-A_B)^{(1/4)} \epsilon ^{(1/4)}$}
\end{deluxetable*}

\end{document}